\documentstyle[12pt,epsfig]{article}

\newcommand{\be}{\begin{equation}}
\newcommand{\ee}{\end{equation}}
\newcommand{\ba}{\begin{eqnarray}}
\newcommand{\ea}{\end{eqnarray}}

\newcommand{\halfs}[1]{\raisebox{-3mm}[0pt][0pt]{#1}}

\begin{document}

\begin{flushright}
{\bf FIAN/TD-03/2000}\\
{\bf BNL-NT-00/13}\\
\end{flushright}

\bigskip

\bigskip

\begin{center}
{\bf AZIMUTHAL ASYMMETRY IN TRANSVERSE ENERGY FLOW IN NUCLEAR
 COLLISIONS AT HIGH ENERGIES}
\end{center}

\medskip

\begin{center}
{\bf Andrei Leonidov$^{(a),(b)}$ and Dmitry Ostrovsky$^{(b)}$}
\end{center}

\medskip

\begin{center}
{\it $(a)$ Physics Department, Brookhaven National Laboratory\\
Upton, NY 11973, USA}
\end{center}

\begin{center}
{\it $(b)$ Theoretical Physics Department, P.N.~Lebedev Physics Institute \\
117924 Leninsky pr. 53, Moscow, Russia \footnote{Permanent address}}
\end{center}

\medskip

\medskip

\begin{center}
{\bf Abstract}
\end{center}

The azimuthal pattern of transverse energy flow in nuclear
collisions at RHIC  and LHC energies is considered.
We show that the probability distribution of the event-by-event
azimuthal disbalance in transverse energy flow  is essentially
sensitive to the presence of the semihard minijet component.

\newpage

At the current level of understanding of the strong interaction physics,
any attempt to give a quantitative description of the bulk of the events
in high energy hadron and nuclear collisions necessarily invokes
several dynamical sources of inelasticity (transverse energy
production), see e.g. \cite{SZ}.
 When materialized in the form of a Monte-Carlo event generator such as
PYTHIA \cite{PYTHIA} or HIJING \cite{HIJING}, the description of
primordial production of partonic degrees of freedom, quarks and gluons,
and of their subsequent conversion into observable hadrons is necessarily
model-dependent. This refers in particular to specific assumptions made
about the contribution of multiple parton collisions, description of the
final state and especially initial state radiation, etc. At the same time
the presence of new physics brought in by semihard degrees of freedom
should manifest itself through reasonably well-defined changes in the
inelasticity
pattern that can (hopefully) be measured experimentally. Ideally the set
of such specific measurements should allow to discriminate between
various otherwise successful approaches (e.g. minijet physics vs.
dual parton model).  Extensive discussion of the relevance of semihard
degrees of freedom (minijets) in explaining such features of high energy
hadron collisions as the enhanced tails of the multiplicity distributions,
multiplicity dependence of the mean transverse energy, etc. by exploiting a
realistic Monte-Carlo event generator HIJING can be found in \cite{HIJING}
and in the review \cite{XNW1}.

 One particular aspect, discussed in \cite{XNW2}, is a $p_{\perp}$
dependence of the two-particle correlation function due to
the semihard minijet contribution in $p {\bar p}$ collisions.
 The  detailed analysis of the minijet contribution to the
two-particle correlation function in heavy ion collisions will appear
in the forthcoming publication \cite{LOL}.
 Below a related calorimetric measure of
the specific event-by-event  azimuthal correlations in the  transverse energy
flow due to minijet contribution will be discussed. The consideration closely
follows the analysis performed in \cite{LO1}, where a specific model of minijet
dynamics based on leading twist collinearly factorized QCD
was considered. The main goal of the present analysis is, by 
using the HIJING event generator,  include the mechanisms
beyond the simple picture of binary parton-parton collisions considered in
 \cite{LO1} such as multiple parton collisions, initial and final state
radiation accompanying these collisions, soft contributions to transverse
energy due to hadronization, jet quenching in nuclear collisions, etc.
Let us note, that accounting for these mechanisms is crucial for
reproducing the transverse energy spectrum \cite{L} 

 In the context of ultrarelativistic heavy ion collisions the primordial
system of semihard degrees of freedom (minijets) is  setting a stage for
subsequent (possibly collective) evolution \cite{BM}. Different approaches
are being used to study the properties of the initial conditions in high
energy nuclear collisions. The conceptually simplest one is based
on accounting for lowest order perturbative contributions in collinearly
factorized QCD, see e.g. the recent papers \cite{CF} and references therein.
Another approach, based on the quasiclassical treatment of the
gluon fields in the colliding nuclei, was considered in a number of
publications \cite{QCL}. A program of the non-perturbative
analysis of primordial gluon distributions is being realized in refs.
\cite{KV}.
Finally, one can rely on the description of minijet effects 
as implemented in the realistic
Monte-Carlo generators such as PYTHIA and HIJING, where effects of
all orders in perturbation theory are (effectively) taken into account.
  In particular, calculations with HIJING allow to study
the effects due to the presence of semihard degrees of freedom
at the early stages of high energy collision in a simple setting, where the
only nontrivial effect distinguishing the nuclear collision from an incoherent
superposition of nucleon-nucleon ones is jet quenching, i.e. energy losses
experienced by partons traversing the surrounding debris created in nuclear
collision.

\bigskip

 To quantify the  event-by-event asymmetry of transverse energy flow, we
consider, following \cite{LO1}, the  difference between the transverse energy
deposited, in some rapidity window $y_{min}<y_i<y_{max}$,
into two oppositely azimuthally oriented sectors with a specified
angular opening $\delta \varphi$ each
\footnote{Long-range correlations in the polar (rapidity) energy flow
in heavy ion collisions were recently considered
in \cite{KLM}.}.

\begin{figure}
\vspace*{-10mm}
\begin{center}
 \epsfig{file=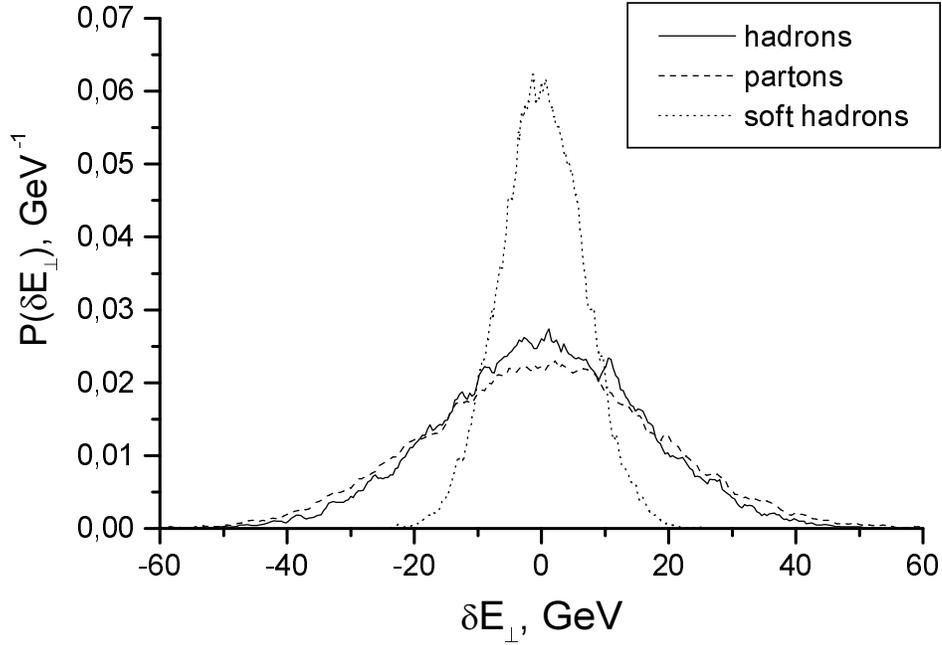,height=9cm,width=13cm}
 \end{center}
\vspace*{-5mm}
 \caption{Probability distribution for azimuthal transverse
energy disbalance in the unit rapidity window for AuAu collisions
at RHIC energy $\sqrt{s}=200\,{\rm GeV}$, $p_0=2\,{\rm GeV}$,
quenching on.}
 \label{frhic}
\end{figure}

\begin{figure}
\vspace*{-5mm}
 \begin{center}
 \epsfig{file=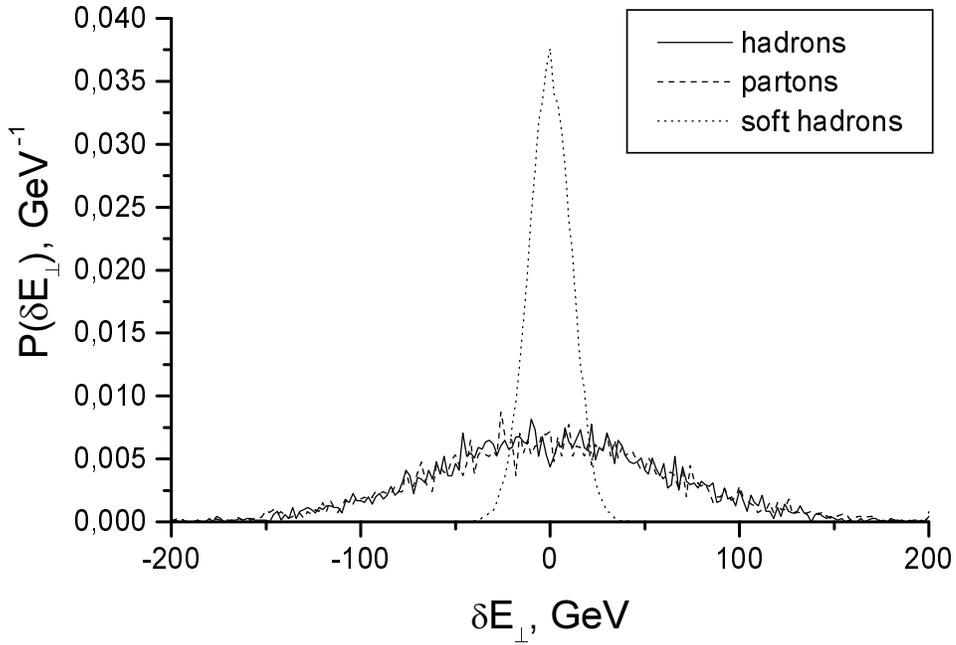,height=9cm,width=13cm}
 \end{center}
\vspace*{-5mm}
 \caption{Probability distribution for azimuthal transverse
energy disbalance in the unit rapidity window for PbPb collisions
at LHC energy $\sqrt{s}=5.5\,{\rm TeV}$, $p_0=2\,{\rm GeV}$,
quenching on. }
 \label{flhc}
\end{figure}

For convenience one can think of the directions of these cones
as being  "up" and "down" corresponding to some specific choice of the
orientation of the system of coordinates in the transverse plane.
All results are, of course, insensitive to the particular choice.
Denoting now the transverse energy going into the "upper" and "lower"
cones in a given event by $E_{\perp}^{\uparrow} (\delta \varphi)$
and $E_{\perp}^{\downarrow}(\delta \varphi)$
correspondingly, we can quantify  
the magnitude of the asymmetry in transverse energy
production in a given event by 
\be\label{deltaE}
\delta E_{\perp} (\delta \varphi) \, = \,
 E_{\perp}^{\uparrow} (\delta \varphi) -
 E_{\perp}^{\downarrow} (\delta \varphi),
\ee
its statistical properties characterized by the corresponding probability
distribution
\be\label{distr}
 P(\delta E_{\perp}|{\delta \varphi}) \, = \,
{d\,w(\delta E_{\perp} (\delta \varphi)) \over
d\,\delta E_{\perp} (\delta \varphi)}
\ee
 We have calculated this distribution in HIJING for central AuAu collisions
at RHIC energy $\sqrt{s}=200\,{\rm GeV}$ and central PbPb collisions at LHC
energy $\sqrt{s}=5.5\,{\rm TeV}$ for $\delta \varphi=\pi$.
The distributions $P(\delta E_\perp|\pi)$ have been
calculated both at partonic level and at the level of final
hadrons with semihard interactions and quenching on and off.
This allowed us to study the contribution of HIJING minijets and of the
effects of their hadronization to the asymmetry in question.
The resulting distributions are plotted in
 Figs.~\ref{frhic} and \ref{flhc}, for RHIC and LHC energies respectively
with quenching turned on and the value of the minijet's infrared cutoff
$p_0=2\,{\rm GeV}$ .

The numerical values of the mean square deviation $\delta E_{\perp}$
characterizing the widths of the corresponding probability distributions
in Figs.~\ref{frhic} and \ref{flhc} are given in Table 1,
where for completeness we also give the widths for the probability
distributions with quenching turned off and with a larger
value for the infrared cutoff $p_0=4\,{\rm GeV}$
\begin{center}
\begin{tabular}{|c|c|c|l|c|}
\hline
AA&$\sqrt{S}$, GeV&$p_0$, GeV & asymmetry &
$\sqrt{\left<\delta E^2\right>}$\strut\\
\hline
            &            &         & hadrons (quenching on)     & 16\\
\cline{4-5}
\halfs{AuAu}&\halfs{200} &\halfs{2}& hadrons (quenching off)    & 17\\
\cline{4-5}
            &            &         & partons                    & 18\\
\cline{4-5}
            &            &         & soft hadrons               &  7\\
\hline
            &            &         & hadrons (quenching on)     & 61\\
\cline{4-5}
\halfs{PbPb}&\halfs{5500}&\halfs{2}& hadrons (quenching off)    & 71\\
\cline{4-5}
            &            &         & partons                    & 65\\
\cline{4-5}
            &            &         & soft hadrons               & 15\\
\hline
            &            &         & hadrons (quenching on)     & 69\\
\cline{4-5}
PbPb        & 5500       &        4& partons                    & 76\\
\cline{4-5}
            &            &         & soft hadrons               & 16\\
\hline
\end{tabular}\nopagebreak

\bigskip
{\bf Table 1}
\end{center}

 The main conclusions that can be drawn from Figs.~\ref{frhic} and
\ref{flhc} and Table 1 are the following.

 First, the magnitude of the azimuthal asymmetry as measured
by the width of the probability distribution
$P(\delta E_{\perp}|\delta \varphi)=
dw(\delta E_{\perp} (\delta \varphi))/d\delta E_{\perp}(\delta \varphi)$ is
essentially sensitive to semihard interactions (minijets).
 Switching off minijets, and thus restricting oneself
to purely soft mechanisms, leads to a substantial narrowing of the asymmetry
distribution; by the factor of 2.3 at RHIC and by the factor 4.1  at LHC
energy rspectively (these values correspond to the case of
quenching being turned on).

Second, quite remarcably, the parton and final (hadronic) distributions
of $\delta E_{\perp}$ in both
cases practically coincide indicating that the contribution to transverse
energy due to hadronization of the initial parton system is, with a high
accuracy,  additive and symmetric in between the oppositely oriented cones.
Both  conclusions show that the energy-energy correlation in
Eq.~(\ref{deltaE})  is a sensitive  measure of the primordial
parton dynamics that can be studied in calorimetric measurements in central
detectors at RHIC and LHC.

Third, as expected, turning off quenching somewhat enhances the fluctuations.
However, as seen from the table, numerically the effect is not important.
This shows once again that the proposed asymmetry is really essentially
determined by the earliest stage of the collision, when the primordial
parton flux is formed.

Finally, from Table 1 we conclude that the studied asymmetry
is not particularly sensitive to changing the value of the
infrared cutoff $p_0$ and thus provides a robust signal of the
presence of semihard dynamics.

\begin{center}
{\it Acnowledgements}
\end{center}

A.L. thanks J.-Y. Ollitrault for useful discussions.

A.L. is grateful to Laboratoire de Physique Theoriquie, Saclay,
where this work was started, and  BNL nuclear theory group, where
it was completed,  for kind hospitality. This manuscript has been authored
under Contract No. DE-AC02-98CH10886 with the U.S. Department of
Energy.

The work was partially supported by  RFBR grant 00-02-16101


\begin{thebibliography}{99}


\bibitem{SZ}
T.~Sjostrand and M.~van der Zijl, {\it Phys.\ Rev.}\ {\bf D36} (1987), 2019

\bibitem{PYTHIA}
T.~Sjostrand, {\it Comput.\ Phys.\ Commun.}\ {\bf 82} (1994), 74;

\bibitem{HIJING}
X.-N.~Wang and M.~Guylassy, {\it Phys.\ Rev.}\ {\bf D44} (1991), 3501;
{\bf D45} (1992), 844;\\
{\it Comput.\ Phys.\ Commun.}\ {\bf 83} (1994), 307

\bibitem{XNW1}
 X.-N. Wang, {\it Phys.\ Repts}\ {\bf 280} (1997), 287

\bibitem{XNW2}
X.-N.~Wang, {\it Phys.\ Rev.}\ {\bf D46} (1992), R1900;
\, {\bf D47} (1993), 2754

\bibitem{LOL}
A.~Leonidov and J.-Y.~Ollitrault "On Azimuthal Correlations in Heavy Ion\\
Collisions", in preparation

\bibitem{LO1}
A.~Leonidov and D.~Ostrovsky, "Angular Pattern of Minijet Transverse
 Energy Flow in Hadron and Nuclear Collisions", [hep-ph/9812416]

\bibitem{L}
A.~Leonidov, ``On Transverse Energy Production in Hadron
Collisions'', {\bf BNL-NT-00/12}

\bibitem{BM}
 J.P.~Blaizot and A.H.~Mueller, {\it Nucl.\ Phys.}\ {\bf B289}
(1987), 847

\bibitem{CF}
K.J.~Eskola, K.~Kajantie, P.V.~Ruuskanen and K.~Tuominen, {\it Nucl.\ Phys.}\
{\bf B570} (2000), 379;\\
K.J.~Eskola, K.~Kajantie and P.V.~Ruuskanen, {\it Eur.\ Phys.\ J.}\ {\bf C1}
(1998), 627;\\

\bibitem{QCL}
A.~Kovner, L.~McLerran and H.~Weigert, {\it Phys.\ Rev.}\ {\bf D52}
(1995), 3809;\\
{\it Phys.\ Rev.}\ {\bf D52} (1995), 6231;\\
Yu.~V.Kovchegov and D.H.~Rischke, {\it Phys.\ Rev.}\ {\bf C56} (1997), 1084\\
S.G.~Matinyan, B.~Mueller and D.H.~Rischke, {\it Phys.\ Rev.}\ {\bf C56}
(1997), 1927;\\
{\it Phys.\ Rev.}\ {\bf C57} (1998), 2197;\\
M.~Gyulassy and L.~McLerran, {\it Phys.\ Rev.}\ {\bf C56} (1997), 2219

\bibitem{KV}
 A.~Krasnitz and R.~Venugopalan, {\it Nucl.\ Phys.}\
{\bf B557} (1999), 237; \\
"Real time simulations of high energy nuclear collisions",
[hep-ph/9808322];\\
"Making Glue in High-Energy Nuclear Collisions", [hep-ph/9905319];\\
"The Initial Energy Density of Gluons Produced in Very High Energy
Nuclear\\
 Collisions", [hep-ph/9909203];\\
"The First Fermi in a High Energy Nuclear Collision", [hep-ph/9910391];\\
"Nonperturbative Gluodynamics of High Energy Heavy-Ion Collisions",
\hbox{[hep-ph/0004116]}

\bibitem{KLM}
Yu.V.~Kovchegov, E.~Levin and L.~McLerran, "Large Scale Rapidity Correlations
in Heavy Ion Collisions", [hep-ph/9912367]

\end{thebibliography}
\end{document}